\documentclass[prd,twocolumn,showpacs,preprintnumbers,amsmath,amssymb,superscriptaddress,floatfix,nofootinbib]{revtex4}

\usepackage{graphicx}
\usepackage{amsmath}
\usepackage{amsfonts}
\usepackage{amssymb}
\usepackage{color}

\usepackage[colorlinks, citecolor=blue,anchorcolor=red,menucolor=red, linkcolor=red,filecolor=red,runcolor=red,urlcolor=blue,frenchlinks=red]{hyperref}

\begin{document}
\arraycolsep1.5pt

\title{The $X(4140)$ and $X(4160)$ resonances in the  $e^+e^-\to \gamma J/\psi \phi $ reaction}

\author{En Wang}
\affiliation{School of Physics and Engineering, Zhengzhou University, Zhengzhou, Henan 450001, China}

\author{Ju-Jun Xie}
\affiliation{Institute of Modern Physics, Chinese Academy of
Sciences, Lanzhou 730000, China}

\author{Li-Sheng Geng}\email{lisheng.geng@buaa.edu.cn}
\affiliation{School of Physics and Nuclear Energy Engineering and
International Research Center for Nuclei and Particles in the
Cosmos, Beihang University, Beijing 100191, China}

\author{Eulogio Oset}
\affiliation{Departamento de
F\'{\i}sica Te\'orica and IFIC, Centro Mixto Universidad de
Valencia-CSIC Institutos de Investigaci\'on de Paterna, Aptdo.
22085, 46071 Valencia, Spain}

\date{\today}

\begin{abstract}
We investigate the $J/\psi \phi$ invariant mass distribution of the  $e^+e^-\to \gamma J/\psi\phi$ reaction at a center of mass energy of $\sqrt{s}=4.6$~GeV measured by the
BESIII collaboration, who concludes that no significant signals are observed for $e^+e^- \to \gamma X(4140)$ because of the low statistics. We show,
however, that the three bump structures in the $J/\psi \phi$ invariant mass distribution, though given only by three data points,  are compatible with
the existence of the $X(4140)$ state, appearing as a peak, and a strong cusp structure at the
$D^*_s\bar{D}^*_s$ threshold, resulting from the  molecular nature of the $X(4160)$ state, which also provides a
substantial contribution to the reaction.  This is consistent with
our previous analysis on the $B^+\to J/\psi\phi K^+$ decay measured by the LHCb collaboration.
 We strongly call for the measurement of this process with more statistics to further pin down the nature of the $X(4140)$ and $X(4160)$ resonances.
\end{abstract}

\maketitle

\section{Introduction}
\label{sec:introduction}
In the last two decades, many  charmonium-like states were observed experimentally, and there exist different  theoretical interpretations on the nature of those states, such as molecular, hybrid, multi-quark states, threshold enhancements. One of the most popular theoretical interpretations is
that they are molecular like, because of the closeness of open coupled channels, see, e.g.,  Refs.~\cite{Chen:2016qju,Guo:2017jvc}.
It is not always easy to firmly identify some states as of molecular nature because of  the existence of other interpretations, such as standard $q\bar{q}(qqq)$ or multiquark states~\cite{Chen:2016qju,Karliner:2017qhf}.

One of the defining features associated to the molecular states that couple to several hadron-hadron channels is that one can find a strong and unexpected cusp in one of the weakly coupled channels at  the threshold of the channels corresponding to the main component of the molecular state ~\cite{Wang:2017mrt,Dai:2018tgo}.

A recent example of this feature is found in the $B^+\to J/\psi\phi K^+$ decay measured by the LHCb collaboration~\cite{Aaij:2016nsc,Aaij:2016iza}, where the analyses, including only the $X(4140)$ resonance at low $J/\psi\phi$ invariant masses, result in a width for the $X(4140)$ resonance much  larger than the average of the PDG~\cite{PDG2016}.  We have investigated this reaction, taking into account the molecular state $X(4160)$, in addition to the $X(4140)$ resonance, and provided a better description of the low $J/\psi\phi$ mass distribution~\cite{Wang:2017mrt}. As predicted in Ref.~\cite{Molina:2009ct}, the $X(4160)$ state is a $D^*_s\bar{D}^*_s$ state with $I^G(J^{PC})=0^+(2^{++})$ and couples to $J/\psi\phi$. As a result the $J/\psi\phi$ mass spectrum develops a strong cusp at the $D^*_s\bar{D}^*_s$ threshold.

Recently, the BESIII collaboration has performed a search for the charmonium-like state $X(4140)$ in the $e^+e^-\to \gamma X(4140) \to \gamma J/\psi\phi$ process at a center of mass (c.m.) energy of $\sqrt{s}=4.6$~GeV, and concludes that no significant signals are observed for $e^+e^-\to \gamma X(4140)$ because of the low statistics~\cite{Ablikim:2017cbv}. However, looking at the $J\psi\phi$ mass distribution of the process $e^+e^-\to \gamma X(4140) \to \gamma J/\psi\phi$ shown in Fig.~5 of Ref.~\cite{Ablikim:2017cbv},  we see that the $J/\psi \phi$ mass distribution exhibits three bump structures around 4135~MeV, 4160~MeV, and 4230~MeV, respectively, given by only three data points,  compatible with
 the peak of the $X(4140)$, the bump at 4160~MeV, and the cusp around the $D^*_s\bar{D}^*_s$ threshold found in the $J/\psi\phi$ mass distribution of the $B^+\to J/\psi\phi K^+$ decay~\cite{Wang:2017mrt}.

In this work, we will analyze the process $e^+e^-\to  \gamma J/\psi\phi$ at $\sqrt{s}=4.6$~GeV, by taking into account the contributions of the $X(4140)$ state, and also the $D^*_s\bar{D}^*_s$ molecular state $X(4160)$. As a test of our interpretation, we also predict the $D^*_s\bar{D}^*_s$ mass distribution of the process $ e^+e^-\to  \gamma D^*_s\bar{D}^*_s$.

This paper is organized as follows. In Sec.~\ref{sec:formalisms}, we present the mechanisms of the  $J/\psi\phi$ and $D^*_s\bar{D}^*_s$ productions in the  $e^+e^-\to  \gamma J/\psi\phi$ and $ e^+e^-\to  \gamma D^*_s\bar{D}^*_s$ processes, respectively, our results and discussions are given in Sec.~\ref{sec:results}. Finally, a short summary is given in Sec.~\ref{sec:summary}.
\section{Formalism}
\label{sec:formalisms}

\subsection{The $J/\psi\phi$ production mechanism}

\begin{figure}
\includegraphics[width=0.35\textwidth]{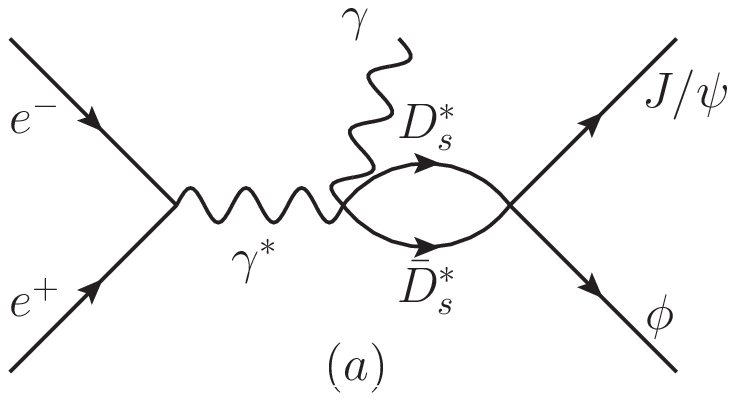}
\includegraphics[width=0.35\textwidth]{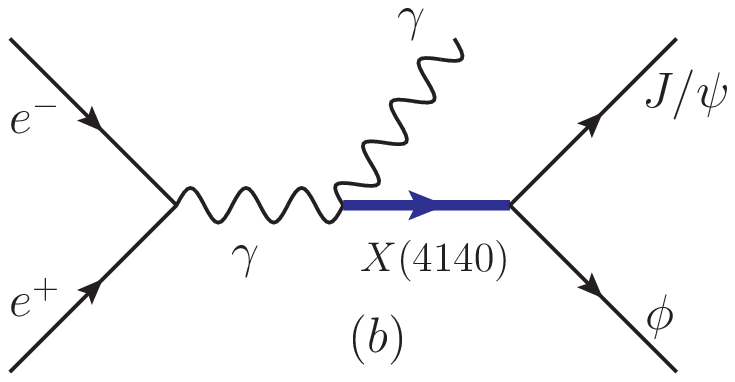}
\caption{Feynman diagrams for the $e^+e^- \to \gamma J/\psi\phi$  process. (a) the contribution of the $D^*_s\bar{D}^*_s$ molecular state $X(4160)$, (b) the contribution of the $X(4140)$ resonance.}  \label{fig:feyn_jpsi}
\end{figure}

In Ref.~\cite{Wang:2017mrt},  the $J/\psi\phi$ invariant mass distribution of the $B^+\to J/\psi\phi K^+$ reaction from threshold to about 4250~MeV~\cite{Aaij:2016nsc,Aaij:2016iza} are described by considering the contributions of the $X(4140)$ and $X(4160)$ states, which tells that both the $X(4140)$ and $X(4160)$ play an important role
in the $J/\psi\phi$ production. Thus, in present work, we extend the mechanism developed in Ref.~\cite{Wang:2017mrt} to the following  $J/\psi\phi$ production reaction,
\begin{eqnarray}
e^- +e^- &\to & \gamma^*(k,\epsilon_1) \nonumber \\
&\to& \gamma(k',\epsilon_2) + J/\psi(p_{J/\psi},\epsilon_{J/\psi}) + \phi (p_\phi,\epsilon_\phi),
\end{eqnarray}
where $\epsilon_1$ ($k$), $\epsilon_2$($k'$), $\epsilon_{J/\psi}$($p_{J/\psi}$), and $\epsilon_\phi$($p_\phi$) are the polarization vectors (momenta) for the virtual photon $\gamma^*$, outgoing photon $\gamma$, $J/\psi$, and $\phi$, respectively. Since the BESIII collaboration measured the process $e^+e^- \to \gamma \phi J/\psi$ at $\sqrt{s}=4.6$~GeV,
the energy of the virtual photon is $k_0\equiv \sqrt{s}=4.6$~GeV.
The Feynman diagrams for the process are depicted in Fig.~\ref{fig:feyn_jpsi}. This reaction can proceed through the $D^*_s\bar{D}^*_s$ interaction, which will dynamically generate the $X(4160)$ resonance, as depicted in Fig.~\ref{fig:feyn_jpsi}(a), and also the intermediate $X(4140)$ resonance of Fig.~\ref{fig:feyn_jpsi}(b). Obviously in the neighborhood of the $X(4140)$ and $X(4160)$ resonances, the tree level term, proportional to the phase space, is small compared to the resonance terms, and therefore we neglect the contribution of the tree level term.

Since the $X(4160)$ state, as a $D^*_s\bar{D}^*_s$ molecular state, has the quantum numbers of $J^{PC}=2^{++}$~\cite{Molina:2009ct}, the vertices of the $X(4160)$ (or the $D^*_s\bar{D}^*_s$ system) to $\gamma^*\gamma $ and $J/\psi\phi$ of Fig.~\ref{fig:feyn_jpsi}(a) are,
\begin{eqnarray}
&&\frac{1}{2}\left(\epsilon_{1i}\epsilon_{2j}+\epsilon_{1j}\epsilon_{2i}\right)-\frac{1}{3}\vec{\epsilon}_1 \cdot \vec{\epsilon}_2 \delta_{ij}, \\
&{\rm and~} &\frac{1}{2}\left(\epsilon_{\phi i}\epsilon_{J/\psi j}+\epsilon_{\phi j}\epsilon_{J/\psi i}\right)-\frac{1}{3}\vec{\epsilon}_\phi \cdot \vec{\epsilon}_{J/\psi} \delta_{ij},
\end{eqnarray}
where the polarizations are evaluated in the rest frame of $J/\psi \phi$,
where the rest of the amplitude is also evaluated.

Thus, the amplitude for Fig.~1(a) is,
\begin{eqnarray}
\tilde{\mathcal{M}}_{J/\psi \phi}^{(a)}&=&A\times G_{D_s^*\bar{D}_s^*} t_{D_s^*\bar{D}_s^*,J/\psi\phi}\times\mathcal{P}^{(a)}  \nonumber \\
&=& \mathcal{M}_{J/\psi \phi}^{(a)}\times\mathcal{P}^{(a)},  \label{eq:amp_Jpsi_FSI}
\end{eqnarray}
with
\begin{eqnarray}
\mathcal{P}^{(a)} &=& \left[\frac{1}{2}\left(\epsilon_{1i}\epsilon_{2j}+\epsilon_{1j}\epsilon_{2i}\right)-\frac{1}{3}\vec{\epsilon}_1 \cdot \vec{\epsilon}_2 \delta_{ij}\right] \nonumber \\
 && \times \left[\frac{1}{2}\left(\epsilon_{\phi i}\epsilon_{J/\psi j}+\epsilon_{\phi j}\epsilon_{J/\psi i}\right)-\frac{1}{3}\vec{\epsilon}_\phi \cdot \vec{\epsilon}_{J/\psi} \delta_{ij}\right],
\end{eqnarray}
where $A$ is an overall normalization factor,  $G$ is the loop function for the intermediate $D^*_s \bar{D}^*_s$ from where by rescattering the final state $J/\psi\phi$ is produced, and $t$ is the transition amplitude of $D^*_s \bar{D}^*_s$ to $J/\psi\phi$. Both $G$ and $t$ depend on the invariant mass $M_{\rm inv}(J/\psi\phi)$.
$\mathcal{P}^{(a)}$ is the spin projection operator, accounting for the spin $S=2$ of the $X(4160)$ state.

For the $G$ function appearing in Eq.~(\ref{eq:amp_Jpsi_FSI}), to avoid potential dangers using the dimensional regularization as pointed out in Ref.~\cite{Wu:2010rv}, we use the cut off method with  fixed $q_{\rm max}=630$~MeV such as to give the same value as $G$ with the dimensional regularization used in Ref.~\cite{Molina:2009ct} at the pole position. With the couplings of $X(4160)(\equiv X_1)$ to $g_{D^*_s\bar{D}^*_s}$ and $J/\psi\phi$ obtained in Ref.~\cite{Molina:2009ct}, the amplitude for the $D^*_s\bar{D}^*_s \to J/\psi \phi$ transition has the following form,
\begin{equation}
t_{D^*_s\bar{D}^*_s,J/\psi \phi} = \frac{g_{D^*_s\bar{D}^*_s}g_{J/\psi \phi}}{M^2_{\rm inv}(J/\psi \phi)-M^2_{X_1}+i\Gamma_{X_1}M_{X_1}}, \label{eq:amp_t1}
\end{equation}
where $g_{D^*_s\bar{D}^*_s}=(18927-5524i)$~MeV  and $g_{J/\psi \phi}=(-2617-5151i)$~MeV, $M_{X_1}=4160$~MeV~\cite{Molina:2009ct}, and,
\begin{equation}
\Gamma_{X_1}=\Gamma_0 + \Gamma_{J/\psi\phi} +\Gamma_{D_s^*\bar{D}_s^*},\label{eq:wid_4160}
\end{equation}
with $\Gamma_0$ accounting for the channels of Ref.~\cite{Molina:2009ct} not explicitly considered here, and,
\begin{eqnarray}
\Gamma_{J/\psi\phi} &=&\frac{|g_{J/\psi\phi}|^2}{8\pi M^2_{\rm inv}(J/\psi\phi)}\tilde{p}_{\phi},\label{eq:dwJpsiphi} \\
\Gamma_{D_s^*\bar{D}_s^*} &=&\frac{|g_{D_s^*\bar{D}_s^*}|^2}{8\pi M^2_{\rm inv}(J/\psi\phi)}\tilde{p}_{D_s^*}\Theta(M_{\rm inv}(J/\psi\phi)-2M_{D_s^*}), \label{eq:dwDsDs} \nonumber \\
\end{eqnarray}
where $\tilde{p}_\phi$ and $\tilde{p}_{D^*_s}$ are the $\phi$ and $D^*_s$ momenta in the rest frame of $J/\psi\phi$ and $D^*_s\bar{D}^*_s$, respectively,
\begin{eqnarray}
\tilde{p}_\phi &=& \frac{\lambda^{1/2}(M^2_{\rm inv}(J/\psi\phi),m^2_{J/\psi},m^2_\phi)}{2M_{\rm inv}(J/\psi\phi)} \\
\tilde{p}_{D^*_s} &=& \frac{\lambda^{1/2}(M^2_{\rm inv}(J/\psi\phi),m^2_{D^*_s},m^2_{D^*_s})}{2M_{\rm inv}(J/\psi\phi)}, \label{eq:mom_ds}
\end{eqnarray}
with $\lambda(x,y,z)=x^2+y^2+z^2-2xy-2yz-2xz$.

In addition, the $J/\psi\phi$ can also be produced via the   $X(4140)$ ($1^{++}$) resonance, as depicted in Fig.~\ref{fig:feyn_jpsi}(b). Taking the suitable operators accounting for the vertex of the  $X(4140)$ [$\equiv X_2$]  to $\gamma^*\gamma$
\begin{equation}
\left( \vec\epsilon_1 \times \vec{\epsilon}_2 \right) \cdot \vec{\epsilon}_{X_2},
\end{equation}
and to $J/\psi\phi$,
\begin{equation}
\vec{\epsilon}_{X_2}\cdot \left( \vec{\epsilon}_\phi \times \vec{\epsilon}_{J/\psi} \right),
\end{equation}
we get
\begin{eqnarray}
\mathcal{P}^{(b)} &=& \sum_{\rm pol} \left[\left( \vec\epsilon_1 \times \vec{\epsilon}_2 \right) \cdot \vec{\epsilon}_{X_2}\right] \left[\vec{\epsilon}_{X_2}\cdot \left( \vec{\epsilon}_\phi \times \vec{\epsilon}_{J/\psi} \right) \right] \nonumber \\
&=&\left( \vec\epsilon_1 \times \vec{\epsilon}_2 \right) \cdot \left( \vec{\epsilon}_\phi \times \vec{\epsilon}_{J/\psi} \right),
\end{eqnarray}
with
\begin{eqnarray}
\sum_{\rm pol}(\epsilon_{X_2})_i (\epsilon_{X_2})_j &=& \delta_{ij}.
\end{eqnarray}
Thus, the amplitude for the diagram of  Fig.~\ref{fig:feyn_jpsi}(b) can be written as,
\begin{eqnarray}
\tilde{\mathcal{M}}_{J/\psi\phi}^{(b)}&=& \frac{B M^2_{X_2}}{M^2_{\rm inv}(J/\psi \phi)-M^2_{X_2}+i M_{X_2}\Gamma_{X_2}} \nonumber \\
&&\times \left( \vec\epsilon_1 \times \vec{\epsilon}_2 \right) \cdot \left( \vec{\epsilon}_\phi \times \vec{\epsilon}_{J/\psi} \right)\nonumber \\
&=&\frac{B M^2_{X_2} \times \mathcal{P}^{(b)}}{M^2_{\rm inv}(J/\psi \phi)-M^2_{X_2}+i M_{X_2}\Gamma_{X_2}} \nonumber \\
&=& \mathcal{M}_{J/\psi\phi}^{(b)}\times \mathcal{P}^{(b)} ,
\end{eqnarray}
where $M_{X_2}=4135$~MeV and $\Gamma_{X_2}=19$~MeV, the same as those of Ref.~\cite{Wang:2017mrt}, and $B$ corresponds to the strength of the contribution of the $X(4140)$ resonance term.

In the present work, the only relevant thing is that the two structures $\mathcal{P}^{(a)}$ and $\mathcal{P}^{(b)}$ do not interfere, and there are no momenta involved, unlike in the decay $B^- \to J/\psi \phi K$~\cite{Wang:2017mrt}.

Hence, we have the $J/\psi\phi$ invariant mass distribution for the process $e^+e^-  \to \gamma J/\psi \phi$  at $\sqrt{s}=4.6$~GeV\footnote{One can perform
the spin sums with the square of the projectors $\mathcal{P}^{(a)}$ and $\mathcal{P}^{(b)}$, but
since $A$ and $B$ are fitting parameters one can omit this operation.},
\begin{eqnarray}
\frac{d \Gamma}{d M_{\rm inv}(J/\psi\phi)} &=& \frac{1}{(2\pi)^3}\frac{1}{4s}k'\tilde{p}_\phi \left[|\mathcal{M}_{J/\psi\phi}^{(a)}|^2+ |\mathcal{M}_{J/\psi\phi}^{(b)}|^2\right],\nonumber \\
\end{eqnarray}
where $k'$ is the momentum of the outgoing photon in the c.m. frame of $e^+e^-$,
\begin{eqnarray}
k'&=&\frac{\lambda^{1/2}\left(s,0,M^2_{\rm inv}(J/\psi\phi)\right)}{2\sqrt{s}}.
\end{eqnarray}

It should be pointed out that  we neglect the mechanism of the $J/\psi\phi$ primarily produced from the virtual photon decay, with a $J/\psi\phi$ intermediate state instead of $D^*_s\bar{D}^*_s$ in Fig.~\ref{fig:feyn_jpsi}(a),  which would involve the extra factor $g_{J/\psi\phi}/g_{D^*_s\bar{D}^*_s}$ versus the amplitude of Fig.~\ref{fig:feyn_jpsi}(a), and we expect to provide a small contribution compared with that of Fig.~\ref{fig:feyn_jpsi}(a).

\subsection{$D^*_s\bar{D}^*_s$ production mechanism}

\begin{figure}
\includegraphics[width=0.35\textwidth]{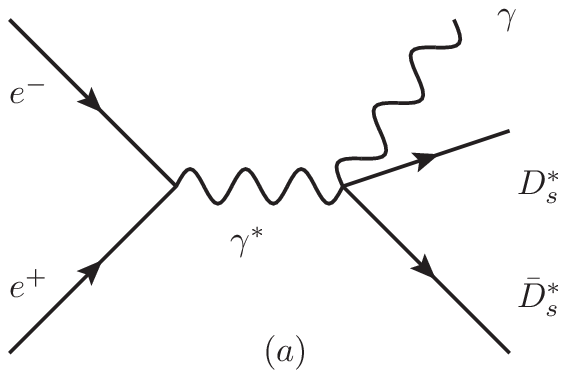}
\includegraphics[width=0.35\textwidth]{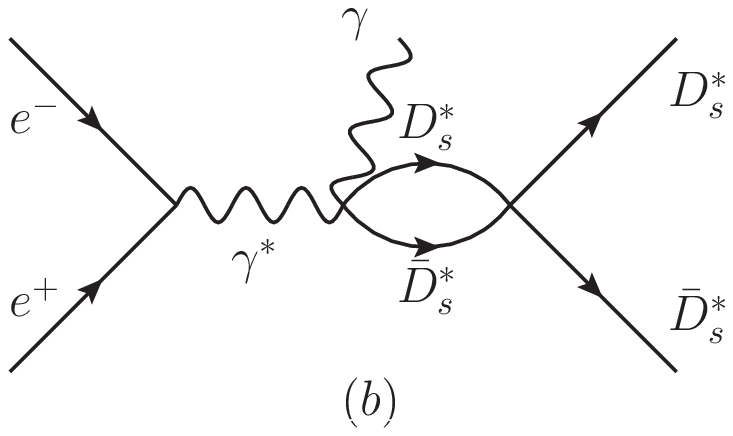}\caption{Feynman diagrams  for the $e^+ e^-\to \gamma D_s^*\bar{D}_s^*$ process. (a) the tree level term,
(b) the contribution of the $D^*_s\bar{D}^*_s$ molecular state $X(4160)$.}  \label{fig:feyn_Ds}
\end{figure}

As the $X(4160)$ state mainly couples to the $D^*_s\bar{D}^*_s$ channel according to Ref.~\cite{Molina:2009ct}, it is interesting to predict the $D^*_s\bar{D}^*_s$  mass distribution of the process $e^+e^-\to \gamma D^*_s\bar{D}^*_s$, which can be used to test the relevance of the $X(4160)$ resonance.

The mechanism of this process is depicted in Fig.~\ref{fig:feyn_Ds}. In addition to the contribution of the $D^*_s\bar{D}^*_s$ molecular state  $X(4160)$, as shown in Fig.~\ref{fig:feyn_Ds}(b), we also take into account the tree level term of Fig.~\ref{fig:feyn_Ds}(a),
which is small compared to the $D^*_s\bar{D}^*_s$ interaction term in the region around the $X(4160)$ resonance. Since the threshold of $D^*_s\bar{D}^*_s$ is about 60~MeV larger than the $X(4160)$ mass, we will keep the tree level term in our calculation.

In analogy to the process of $e^+e^- \to \gamma J/\psi \phi$, the $D^*_s\bar{D}^*_s$ mass distribution of the process $e^+e^-\to \gamma D^*_s\bar{D}^*_s$ at $\sqrt{s}=4.6$ can be written as,
\begin{eqnarray}
\frac{d \Gamma}{d M_{\rm inv}(D^*_s\bar{D}^*_s)} &=& \frac{1}{(2\pi)^3}\frac{1}{4s}k'\tilde{p}_{D^*_s} |\mathcal{M}_{D^*_s\bar{D}^*_s}|^2, \\
\mathcal{M}_{D^*_s\bar{D}^*_s} &=& A \left[ T^{\rm tree}+ T^{X(4160)}\right] \nonumber \\
&=& A \left[1 + G_{D^*_s\bar{D}^*_s}\left(M_{\rm inv}(D^*_s\bar{D}^*_s)\right) \right. \nonumber \\
&&\left. \times t_{D^*_s\bar{D}^*_s,D^*_s\bar{D}^*_s}\left(M_{\rm inv}(D^*_s\bar{D}^*_s)\right)\right],
\end{eqnarray}
where the factor $A$ and the loop function $G$ are the same as those of
Eq.~(\ref{eq:amp_Jpsi_FSI}). The transition amplitude $t_{D^*_s\bar{D}^*_s,D^*_s\bar{D}^*_s}$ is given in terms of the coupling $g_{D^*_s\bar{D}^*_s}$ obtained in Ref.\cite{Molina:2009ct} by,
\begin{equation}
t_{D^*_s\bar{D}^*_s,D^*_s\bar{D}^*_s} = \frac{g^2_{D^*_s\bar{D}^*_s}}{M^2_{\rm inv}(D^*_s\bar{D}^*_s)-M^2_{X_1}+i\Gamma_{X_1}M_{X_1}}.
\end{equation}

\section{Results and discussions}
\label{sec:results}

\begin{figure}
\includegraphics[width=0.4\textwidth]{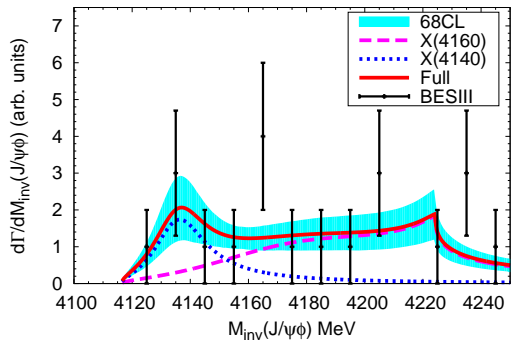}
\caption{The $J/\psi\phi$ invariant mass distribution of the process $e^+e^- \to \gamma J/\psi \phi$. The magenta dashed line and the blue dotted line show the contributions of the $X(4160)$ [Fig.~\ref{fig:feyn_jpsi}(a)] and $X(4140)$ resonances [Fig.~\ref{fig:feyn_jpsi}(b)], respectively, and the red solid line corresponds to the full contribution. The experiment data is taken from the BESIII measurement~\cite{Ablikim:2017cbv}. The band reflects the uncertainties in $A$ and $B$ from the fit, and represents the 68\% confidence-level.} \label{fig:dw_jpsi}
\end{figure}

\begin{figure}
\includegraphics[width=0.4\textwidth]{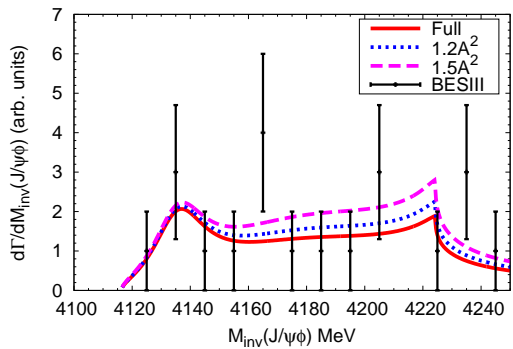}
\caption{The $J/\psi\phi$ invariant mass distribution of the  $e^+e^- \to \gamma J/\psi \phi$ process by increasing $A^2$ to $1.2A^2$ (magenta dashed line)  and $1.5A^2$ (blue dotted line). The other explanations are same as Fig.~\ref{fig:dw_jpsi}. } \label{fig:dw_jpsi2}
\end{figure}
As we discussed above, there are three free parameters in our model for the process $e^+e^- \to \gamma J/\psi \phi$: (I) $A$, an overall normalization factor, (II) $B$, the strength of the contribution of the $X(4140)$ resonance term, (III) $\Gamma_0$, accounting for the channels of Ref.~\cite{Molina:2009ct} not explicitly considered in this paper. In our previous work~\cite{Wang:2017mrt}, a similar mechanism of the $J/\psi\phi$ production is used to describe the $J/\psi\phi$ mass distribution of the  $B^+\to J/\psi\phi K^+$ decay measured by the LHCb collaboration~\cite{Aaij:2016nsc,Aaij:2016iza}, where the $\Gamma_0$ is  extracted to be $67.0\pm 9.4$~MeV. We will take this value of $\Gamma_0$, and fit the other two parameters ($A$ and $B$) to the BESIII data of the $J/\psi\phi$ mass distribution of the process $e^+e^- \to \gamma J/\psi \phi$ from threshold to 4250~MeV at $\sqrt{s}=4.6$~GeV~\cite{Ablikim:2017cbv}. The resulting $\chi^2/d.o.f \sim 6.1/(12-2)=0.61$ is very small, mainly because of the large errors of the data. We present the $J/\psi\phi$ mass distribution in Fig.~\ref{fig:dw_jpsi}, where we can see that there is a significant peak around 4135~MeV, associated to the $X(4140)$ resonance, which is in agreement with the BESIII measurement, although the BESIII collaboration concludes that there is no structure  in the $J/\psi\phi$ mass distribution because of the low statistics.
In addition, we also find a broad bump around the mass of the $X(4160)$ resonance, and a sizeable cusp structure around the $D^*_s\bar{D}^*_s$ threshold, both resulting from the dynamically generated  $X(4160)$ resonance, which is also compatible with the low statistics measurement of the BESIII collaboration~\cite{Ablikim:2017cbv}. 	

It should be noted that the peak around 4135~MeV, and the cusp structure close to the $D^*_s\bar{D}^*_s$ threshold, if confirmed in more accurate measurements, should be associated to the $X(4140)$ and $X(4160)$ resonances.
 In Fig.~\ref{fig:dw_jpsi2}, we present our results by increasing $A^2$, and find that the broad bump and the cusp become more clear. The cusp comes from the factor $G_{D^*_s\bar{D}^*_s}$, and reflects the analytical structure of this function with a discontinuity of the derivative at threshold. The bump and the cusp structure appearing in Fig.~\ref{fig:dw_jpsi} are due to the $D^*_s\bar{D}^*_s$ molecular nature of the $X(4160)$ resonance. Thus, we strongly suggest that the BESIII collaboration measure this process with better statistics.

\begin{figure}
\includegraphics[width=0.4\textwidth]{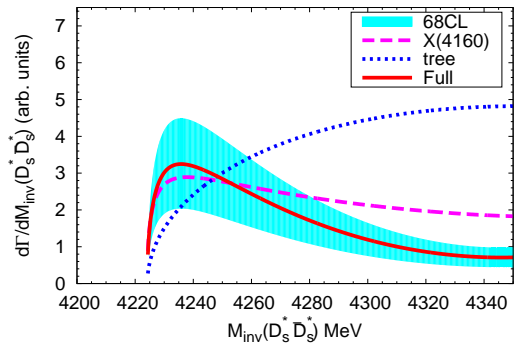}
\caption{The $D^*_s\bar{D}^*_s$ invariant mass distribution of the process $e^+e^- \to \gamma D^*_s\bar{D}^*_s$. The magenta dashed line and the blue dotted line show the contributions of the $X(4160)$ resonance  [Fig.~\ref{fig:feyn_Ds}(a)] and the tree level term [Fig.~\ref{fig:feyn_Ds}(b)], respectively, and the red solid line corresponds to the full contribution. The band reflects the uncertainties  in $A$ and $B$ from the fit, and represents the 68\% confidence-level.} \label{fig:dw_ds}
\end{figure}

Finally, as a test of the interpretation given here, we also present  in Fig.~\ref{fig:dw_ds} the $D^*_s\bar{D}^*_s$ invariant mass distribution for $e^+e^- \to \gamma D^*_s\bar{D}^*_s$, with the parameters fixed above. It is shown that there is an enhancement close to the threshold, significantly different from  the phase space (labeled as 'tree'). The peak is the reflection of the $X(4160)$, and should not be mis-identified as a new resonance.

\section{summary}
\label{sec:summary}
Recently, the BESIII collaboration studied the $J/\psi \phi$ invariant mass distribution of the process  $e^+e^-\to \gamma J/\psi \phi$ at the c.m. energy of $\sqrt{s}=4.6$~GeV, and pointed out that there is no structure because of the low statistics.  However, the three bump structures around 4135~MeV, 4160~MeV, and 4230~MeV, though only given by three data points, are compatible with our previous analyses on the LHCb measurement of the reaction $B^+\to J/\psi\phi K^+$.

In this work, based on our previous work about the decay of $B^+\to J/\psi\phi K^+$, we analysed the process of $e^+e^-\to \gamma J/\psi \phi$ , by considering the contributions of the $X(4160)$ resonance, as a $D^*_s\bar{D}^*_s$ molecular state, and the $X(4140)$ resonance.  Because of the large errors of the BESIII data, the $\chi^2/d.o.f$ of the fit is very small. We found a peak around 4135~MeV, associated to the $X(4140)$ resonance, and a broad bump and a cusp structure, which appear as a consequence of the  $D^*_s\bar{D}^*_s$ molecular structure of the $X(4160)$ resonance. Thus, we strongly call for a measurement of this process with high precision. Finally, as a test of our interpretation, we predicted the  $D^*_s\bar{D}^*_s$ mass distribution of the process $e^+e^- \to \gamma D^*_s \bar{D}^*_s$ at  $\sqrt{s}=4.6$~GeV, and found an enhancement close to the threshold, which is the reflection of the $X(4160)$ resonance and should not be mis-identified as a new resonance.

\section*{Acknowledgements}

This work is partly supported by the National Natural
Science Foundation of China under Grant Nos.11475227,
 11735003, 11522539, 11505158, 11475015, and  11647601.  It is also supported by
the Youth Innovation Promotion Association CAS (No.
2016367),
the Academic Improvement Project of Zhengzhou University, the China Postdoctoral Science Foundation under Grant No. 2015M582197, and the Postdoctoral Research Sponsorship in Henan Province under Grant No. 2015023.
This work is also partly supported by the Spanish Ministerio de Economia
y Competitividad
and European FEDER funds under the contract number FIS2014-57026-REDT,
FIS2014-51948-C2-1-P, and FIS2014-51948-C2-2-P, and the Generalitat
Valenciana in the program Prometeo II-2014/068.

\end{document}